\documentclass[9pt,oneside]{amsart}
\usepackage{url}
\usepackage{xspace}
\usepackage{graphicx}
\usepackage{multicol}
\usepackage{subfig}
\usepackage{amsmath}
\usepackage{amssymb}
\usepackage[a4paper,width=170mm,top=18mm,bottom=22mm,includeheadfoot]{geometry}
\usepackage{booktabs}
\usepackage{array}
\usepackage{verbatim}
\usepackage{caption}
\usepackage{natbib}
\usepackage{float}
\usepackage{pdflscape}

\newcommand*\eg{e.g.\@\xspace}
\newcommand*\Eg{e.g.\@\xspace}
\newcommand*\ie{i.e.\@\xspace}

\title{A C++ Language Workbench}
\author{Gavin Wood}

\begin{document}

\begin{abstract}
Language-orientated programming promises to elevate programmer productivity through increased abstraction capabilities. Structural programming environments provide apparatus to reduce the difficulties with syntax. The language workbench, a conceptual combination of these two approaches, is a comparatively novel approach to software development and has so far been attempted only in dynamic-dispatch, run-time-compiled languages (\eg Java).

However, it must be remembered that several fields of engineering exist, each having their own priorities. In the video games industry, where large, complex and diverse projects are routinely developed, efficiency is paramount and as such C++, as a development platform, is widely used. I explore the possibility of a language workbench capable of a gradual transition in both skills and code from the traditional C++ development environment.

This article is the design for a language workbench. It uses novel techniques including a context-sensitive event-driven input system and a hybrid single/multiple-inherited class model and through a prototype implementation demonstrates that is both concise and practical for C++. I refute the hitherto implicit hypothesis that the language workbench paradigm is not applicable to the C++ language, showing that C++ can be used for creating an effective development framework usable in otherwise pure-C++ programming environments.
\end{abstract}

\maketitle

\setlength{\columnsep}{20pt}
\begin{multicols}{2}

\section{Introduction}\label{sec:introduction}

Despite decades of research programming computers, in general, remains difficult. Incremental improvements to the development stack---\eg our code editors, languages, prototyping and versioning tools---have made larger and more complex systems possible to build. New programming paradigms---\eg language-orientated programming---have allowed our code to become more conceptually powerful, reducing language-induced idioms, improving the ``meaning-to-noise'' ratio and allowing the programmer to focus on creativity over rules.

Some issues of development, however, \eg syntax errors, file/directory management and refactoring will not be fixed by better text-editors, cleverer codes or more auxiliary tools; they are implications of this traditional development stack. To solve these problems we must ultimately move past languages and directories of text files: \textit{language workbenches} are one such destination.

Some industries however, require the combination of complex, high-performance software and languages/development stacks that are well-tested, portable, industrially-supported, compatible with legacy code and for which a large developer base already exists. In this case there is little option; only C++ can fulfil these requirements.

\subsection{Overview}

With the lack of a C++-based language workbench, it might be tempting to conclude that an implementation would be impractical (perhaps due to the lack of language features of C++) or the eventual benefits minimal (perhaps too cumbersome or unwieldy). The present work dispels this notion. Here I outline and discuss a highly flexible and extensible language workbench based around the C++ language.

Core functionality is reduced to a minimum, providing instead flexible mechanisms allowing lingual idioms to be developed modularly. In particular, display and input are fundamentally separated; text-based input is eschewed; the program is guaranteed to be structurally correct throughout editing; and only a simple hierarchical program structure is explicitly provided for. Potentially superfluous features such as program transformations, meta-language programming and cross-references between program entities are by design not part of the core.

In order to implement this efficiently in C++, several issues are addressed in the eventual design: a hybrid single/multiple inheritance schema is used to best express the complex interrelationship of language concepts; an extensive `undo' infrastructure is used to supplant traditional serial-entry-specific mechanisms; a highly extensible display mechanism based around web-technologies is used to provide rich visualisation, interaction and computer-aided development; and a multi-modal hierarchy structure is used to facilitate metaprogramming techniques.

I outline design and implementation issues and present clear solutions to deliver a solution efficiently and concisely in cross-platform C++, comparing where applicable to past and present approaches. I show the prototype's operation demonstrating how the received features are useful. Finally, I detail future directions that the present work, and language workbenches in general, might take.

\section{Background}\label{sec:background}

\subsection{Structural Editing}

Structural editing can be seen as a move away from the text-editor/text-file/compiler triplet that still dominates modern programming environments. The actual idea of editing the conceptual  program model rather than a string of plain text is at least 30 years old. When describing an early program-model editor, \textit{Cornell Program Synthesizer}, \citet{teitelbaum1981} points out that ``programs are not text; they are hierarchical compositions of computational structures and should be edited, executed, and debugged in an environment that consistently acknowledges and reinforces this viewpoint''. The idea did not become popular; indeed one might ponder how many programmers even recognise the difference between the code of a program and its model.

More modern structural editors do exist and recent work into usability aspects has continued; \cite{goldman2004} describes a Java structural editor, \textit{JPie}, used for the instruction of novice programmers. However there is common criticism of structural editors (e.g. discussed by \cite{birnbaum2005}) which relates to the difficulty of editing a program model with serialised input: a program-model that is semantically correct, if entered naturally, requires transition states of the program model that are nonsensical. For example, assigning an integer variable \texttt{i} from the output of a function \texttt{foo()} one might like to enter \texttt{i = foo} before putting the final parentheses on \texttt{foo} to dictate the desire to invoke it rather than assign it as a functor. \cite{ko2004} provides a more circumspect discussion about structural-editing issues found as part of the Natural Programming project (a Java-based repertoire of structural programming tools) and finds that the user-interface of a structural code editor must be carefully designed to be not only supportive (\ie strict) but flexible enough to facilitate the user's manipulation of the program.

\begin{figure*}[htb]
\begin{center}
\label{fig:traditional}
\subfloat[Traditional]{\includegraphics[scale=0.70]{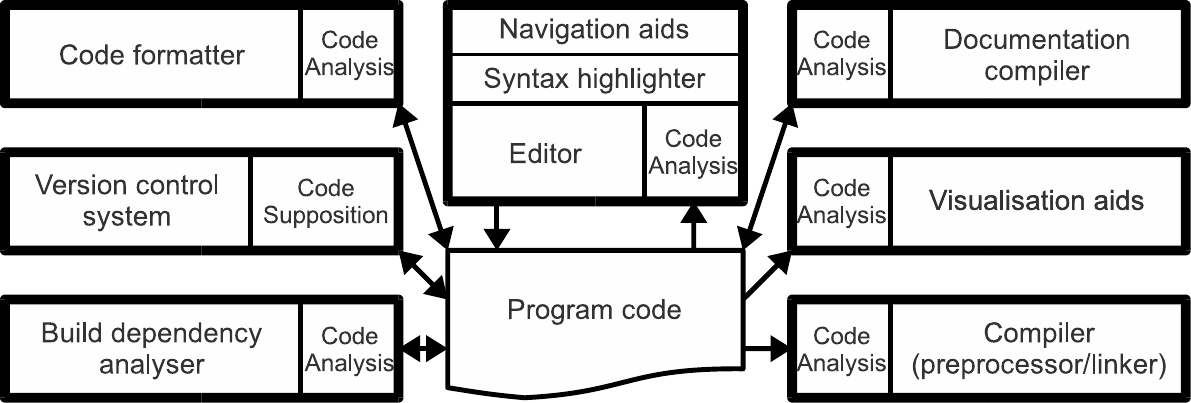}~}
\subfloat[Structural]{~\includegraphics[scale=0.70]{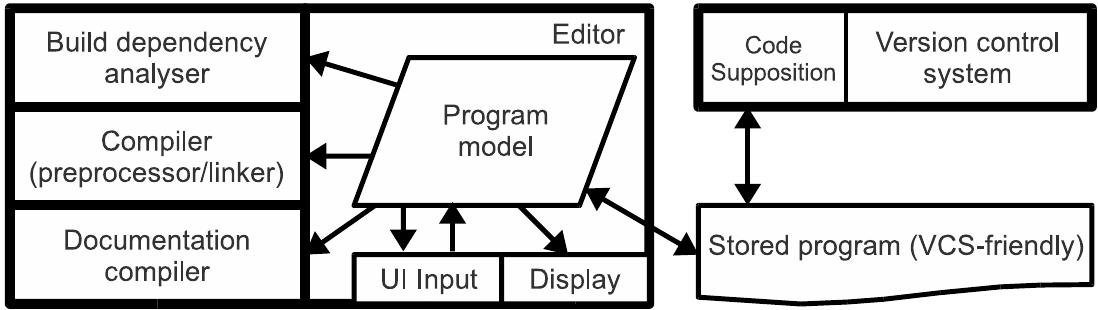}}
\caption{The organisation of a number of components of the development system.}
\end{center}
\end{figure*}

\subsection{Language-Orientated Programming}

\textit{Language-Orientated Programming} (LOP) coined by \cite{ward1994} and discussed later by \cite{dmitriev2005}, brings together such paradigms as intentional programming, model-driven architecture (MDA) and generative programming. Put simply, LOP champions a \textit{middle-out} development order with an abstraction mechanism that contains syntax, grammar and vocabulary.

LOP has several indisputable advantages, as described by \cite{simonyi2006}, including greater scope for parameterisation (thereby optimising code reuse), maximising productivity through facilitating the mixing of multiple languages and providing the canvas of language grammar to better help programmers codify their thoughts. To aid in the continuous development of new languages that LOP requires, a new class of structural editors have come into existence termed \textit{language workbenches}.

\subsection{Language Workbenches}

Combining the currents of structural program editing and language-orientated development, language workbenches provide a development framework that has the potential to alleviate or eliminate traditional problems such as syntax errors, file-management and refactoring, while providing the programmer with a tool flexible enough to program a computer with whatever dialogue (language/paradigm) she sees fit.

A structural (imperative) programming editor is an environment for developing subroutines in terms of other co-developed subroutines. In this manner, a language workbench can be considered an environment for developing languages in terms of other co-developed languages. Put another way, a language workbench is a structural editor, where the abstract grammar (or \textit{program form}) can itself be specified from within the same environment in the very same manner as one would otherwise write program code.

Several language workbenches have been proposed. Chief among them are the graph-based \textit{GReAT} framework (proposed by \cite{agrawal2003}), the Lisp-based \textit{FermaT} program manipulation system (proposed by \cite{ward2005}), JetBrains's open-source Java-based \textit{Meta-Programming System} (MPS; introduced by \cite{dmitriev2005}), and IntentSoft's C\#-based \textit{Domain Workbench} (discussed by \cite{simonyi2006}). All are implemented in imperative or run-time compiled, dynamic-dispatch languages (incompatible with C++) and have either cumbersome input designs or fall back onto a textual language input system.

The \textit{FermaT} program manipulation system and the framework it is built in work on the assumption that an LOP project should comprise a single stack of formally-defined languages, each more specific to the ultimate problem's domain. Thus it becomes trivial to write the ``final'' program in the highest-level language. One downfall of the single-stack design is to make difficult the combination of distinct languages, thus hampering reuse.

The \textit{Meta-Programming System} (MPS), proposed and developed by JetBrains is a now open-source comprehensive language workbench implemented in Java. It is based around a Java-implemented editor and several conceptual languages; here \textit{language} is used to define the structure, editor (\ie I/O) and semantics of a set of concepts. In particular, a language needs to have no canonical textual representation. The languages MPS comes with include a structure---or language---language (\ie one that is defined in terms of itself, but initially would be bootstrapped from an absolute reference implementation); several base languages, corresponding to key portions of the Java platform; several language-transformation languages (allowing operations on abstract portions of a program); and a editor language, allowing new editors to be created for the languages. The display is extensible through the editor language, but a cellular display---not dissimilar from an appropriately customised spreadsheet---is provided as standard. Being Java-based it is ineffective for authoring C++ projects and extending the C++ language.

The C\#-based \textit{Domain Workbench} exists as perhaps the most comprehensive and advanced proposal for a language workbench. It is described as a programming environment centred around the \textit{Intentional Tree}, an abstract data type representing, as best as possible, what it is that the author means. The tree is a description of the program in terms of a number of \textit{def}s; these defs form the \textit{domain schema} and are themselves defined as intent trees. The tree can be \textit{projected} to the user with a number of pre-defined projections (assuming some cooperation is present between the projection and the \textit{def}s that the tree references). The intentional tree may be edited through several ``intuitive'' tree-based operations (splicing, inserting children, appending on a list, \&c.). Domain schemas may be composed to form hybrid schemas. A special \textit{generator} module, directed by the `domain code' (represented by the intentional tree) is able to generate the C\# sources to be compiled with the help of the domain schema. Creating content in the intentional tree is done by entering text, which is parsed and acted upon according to the current domain schema, as such intermediate languages must be designed very carefully to keep the parse-tree unambiguous. Being C\#-based, it is ineffective for working with C++ projects or building intermediate languages on the C++ language.

From this brief review it is clear that the design and implementation of a language workbench based around C++ has not yet been addressed. Furthermore, several deficiencies (such as a reliance on a notional language's text representation for input) are, I argue, both unnecessary and fixable.

\subsection{Terminology}

Terminology in the different language workbenches is fragmented; a quick lookup-table is provided in below. Henceforth, I elect to use the MPS terminology, refining the term \textit{language} to \textit{abstract language} to avoid any mistaken inference that a form of syntax may be involved.

\begin{table}[H]
\label{tab:terms}
\begin{center}
\begin{tabular}{lll}
\toprule
Presently & DW & MPS \\
\midrule
abstract language & domain schema & language \\
concept & def/identity & concept \\
program(-model) & domain code & program \\
node & node & node \\
\bottomrule
\end{tabular}
\end{center}
\caption{Language workbench terms in use.}
\end{table}

As such we may say that a \textit{program} is a structure of \textit{node}s, each of which has an `is-a' relation to a single \textit{concept} composing the overall \textit{abstract language}. Furthermore, any \textit{concept} can be (recursively) defined as a \textit{program}, given the appropriate \textit{abstract language}.

\section{The System}

The open-source \textit{Martta} project (\url{http://martta.net}) is the research vehicle I am using to demonstrate compatibility between traditional C++-development and the language workbench paradigm. It must:

\begin{itemize}
\item be portable to major platforms---Win32 (see \cite{hart2004windows}), Mac OS X (see \cite{hillegass2008cocoa}), Linux (see \cite{peterson1998linux});
\item have an initial abstract language based around the C++ language;
\item support native cooperation with C++ legacy code and the compilers/linkers;
\item be natural to C++ developers in terms of code display and manipulation;
\item support development through avoidance of structural errors and recognition of the existence and nature of static semantic errors;
\item have an extensible display with customisable input mechanisms;
\item be extensible regarding abstract and implementation language grammar (in particular, the ability to utilise C++0x should require a minimal amount of change);
\item be compatible with or adaptable to existing tools such as version control systems.
\end{itemize}

%Ultimately, with the abstraction facilities that an LOP environment provides, the fact it is based around C++ would be of little importance in terms of the design. However, boot-strapping must be conducted once in order to reach this point and such facilities are not available then. Thus the language facilities of C++ were considered well in order to forge a design that can be implemented as swiftly and concisely as possible while remaining compatible with the overall goals.

To maximise the extensibility, a design priority was to make the core components (i.e. those which may not be swapped out at a later stage) as minimal as possible. In particular, the core is as small and flexible as possible so that functionality for C++ language compatibility could be provided, concisely, by extensions. This raises the likelihood that future exotic ideas will be easily implemented avoiding changes to the core. Furthermore, the abstract language must be well extensible without falling back on the basic LOP abstraction mechanism of simply implementing a new language in terms of the old. Consider a hypothetical progression from the C-subset of C++ to the complete C++ abstract language: ideally, the most concise C-subset implementation could be extended to become the most concise full C++ implementation; similarly, it should not be necessary to make changes to this C++ implementation in order to extend it into a concise C++0x implementation.

%\begin{figure}[H]
%\label{fig:system}
%\begin{center}
%\includegraphics[width=6.5cm]{System.pdf}
%\end{center}
%\caption{A general overview of the Martta system.}
%\end{figure}

The system I propose can be split functionally into four distinct portions which have a set of run-time link dependencies:
%. Figure \ref{fig:system} depicts them:

\begin{itemize}
\item Editor: An executable GUI `harness' for utilising the computer's I/O and hosting everything else, including an API for extending its functionality.
\item Concept API: The base class (called \texttt{Concept}) representing the required interface for defining a language concept together with the class's support mechanisms.
\item A number of GUI \textit{extensions} and their associated components.
\item A number of concept classes, forming a primitive abstract language.
\end{itemize}

The first two portions (the editor and base concept class) are essentially fixed and constant parts of the system. They are separate to allow the base concept class to be linked to specialised concept classes without the bulk of the editor. The GUI extensions allow augmentation of the editor (e.g. different input mechanisms, views \&c.) and help specify the user's interface. They have a link-time dependency on the editor. Finally, the concept classes are a set of specialised base concept classes which together define the language in which the user must express their program. This set of classes is independent of the editor though are linked to the editor at run-time.

\subsection{Editor}

The editor forms a core ``harness'' or, as \cite{birsan2005} refers to it, a \textit{plug-in engine}, for each of the other components. Aside from loading editor extensions, the key functionality of the editor is threefold:

Firstly, to manage programs, holding them in memory, providing access to, saving and loading them to disk (trivially accomplished through the concept's model navigation, RTTI-style inspection and the property specification API).

Secondly, to route basic user-interface events such as (multiple) key presses, mouse/touch gestures, menu items \&c. to specific transforms on the program of \textit{actions}, parameterising them according to the current display. It maintains a display of all actions that can be performed immediately in a manner similar to \textit{Cepage}, described by \cite{meyer1988}.

Thirdly, to provide execution facilities. Fundamentally, program semantics are defined through the concept classes of the program. The program's root concept's class defines, through an interface known to the editor, how the program can be executed. Having all semantics information in only a single place presents some difficulties for making the language easily extensible; this is solved by decentralising meaning to individual fragments of the program/abstract language as described in section \ref{sec:semantics}.

\subsection{Concept}

The fundamental base of the Martta system is a class named Concept. All classes derived from Concept are known henceforth as \textit{concepts} and together form the class hierarchy rooted at Concept. An abstract language may be described as a set of such classes. To allow certain simplifications in the design of the present system, a number of restrictions are made on, and facilities provided to, concepts. As they are fundamentally classes, concepts have function and data members (aka properties). \Eg the \texttt{StringLiteral} concept has a datum \textit{text}, which stores the quoted text of the literal.

A program is defined as a strictly hierarchical set of nodes, each node being an instantiation of a particular concept. The concept of a node defines, among other things, the allowed concepts of its child nodes. In this way, an abstract language may simply and effectively define structural-correctness in much the same way a grammar is defined, prohibiting most invalid programs immediately and simplifying both the present project's implementation and the eventual job of the user. Unfortunately, with a strictly single-inheritance model for the concept class hierarchy, this leads to difficulties in code-sharing and program cross-referencing. This was overcome with a hybrid inheritance model, detailed later.

There are some concepts, equivalents of which would not be found in a traditional language's grammar. Perhaps because they deal with aspects of a language that is traditionally implicit (e.g. its display), perhaps because they provide an abstraction that is unimportant in a purely grammatical sense. These concepts often make sense to be considered discretely (perhaps providing appropriate code and hooks), though being grammatically inert and never directly realised as nodes. These are so-called \textit{notional} concepts and are problematic since they imply a structure more complex than a simple hierarchy. The ramifications of these are discussed later.

\subsection{Extensions}

Editor extensions (or simply \textit{extensions}) allow concepts of a language to interact with the GUI editor in novel ways. They take the form of a dynamically loadable library that interfaces with the GUI to extend its functionality, and that interfaces with the concept hierarchy to provide language components to use this GUI functionality. One such example is that of the display:

The GUI editor must have at least one display; \ie a component capable of allowing the user to navigate around a program and maintain a number of properties such as the currently focused node. As an example design, I have implemented such a component, based around a HTML/CSS/JavaScript document, displayed by a WWW page display engine. It includes a `stylist' delegate class (to allow some customisation of the appearance of the program) and a notional concept (\texttt{WebViewable}) for a node that can present itself for such a view.

The view displays the program simply by asking the program's root node for an HTML representation. To form this HTML representation, the node may defer to its children recursively. If a node does not implement the interface for specifying an HTML representation (a concept-class known as \texttt{WebViewable}) then a basically-styled HTML dump of the node is returned as a fallback. JavaScript is used in the definition for basic interactivity and CSS for abstracting component of style. Configuring or reimplementing the \textit{stylist} delegate (which handles the fallback mechanism) allows the display to be customised.

In this way, the actions one performs are fundamentally separate from visual appearance those actions invoke.

\subsection{The Abstract Language}

Initially, the abstract language is the set of concepts loaded into the editor prior to any program. Being C++-based, this system begins with an abstract language conceptually similar to C++.

If, within her project, a programmer does author a new concept, then it augments the abstract language for all independent parts of the project (the editor implicitly compiles the class, links and dynamically loads it). In much the same way as a programmer may specify a library dependency in C++, she may specify a abstract language dependency in Martta (the editor would dynamically load and link the requested concept classes).

Two core Martta abstract languages are proposed; one approximating the grammar of the C++ language, another (based around the \texttt{Concept} class and its ancillaries) providing the necessary functionality for authoring new concepts. The latter includes concepts for program transformation, program fragment literals and parameters for such literals.

\section{Difficulties and Design Solutions}

The system outlined so far, has a number of serious flaws that must be addressed for it to fulfil the requirements given.

\subsection{Node Organisation}
\label{sec:nodeorganisation}

Extensibility is maximised by pushing as much functionality as possible outside the core of Martta and into the specialised concept classes. An example of this is the structure of the program; it is a simple hierarchy of nodes. In particular there is no explicit support for node cross-references (unlike \eg \textit{MPS}). In the present solution, such functionality is contained within the C++ abstract language (so that, for example, an invocation can refer to the function declaration).

\subsubsection{Child Indexing}

However, a hierarchy where each node may have an (unordered) set of child nodes is somewhat restrictive. With most languages being serialised, the order of nodes is typically important: in C++, for example, the argument order of a function call is generally paramount. As such, an ordered set of children for each node seems reasonable. This is better, but still not ideal; many language concepts require their nodes have not only an unbounded ordered list of children, but also a number of other children fulfilling a special role. Consider in C++ the function call; this requires the argument values (as mentioned earlier), but also the value of the invocation function.

Two immediate solutions present themselves; one may persist with the ordered list of children (each child has an integer index into its parent's list), perhaps using offsets and reserved places in the list for maintaining children outside the cardinal order. One may alternatively use an associative-array structure, requiring children associated with the same key to retain their order (each child having an associated key and index into the key's list). Unfortunately both have disadvantages; the first is clumsy and reduces extensibility, the second is inefficient to implement. Rather than either of these two, a hybrid model was used whereby each node has \textit{both} an ordered list of children and an associative array.

As in the first solution, nodes have a single integer index into their parent's set of children. However, this index only refers to a position in an ordered set if it is non-negative. When negative, it refers to the value in the associative array. These two types of indexing method are called respectively \textit{cardinal} and \textit{named}. Actual negative numerals are never actually used; variables are always used for named indices in development \& storage and are mapped to arbitrary negative numbers at compile-time using the in-memory addresses of \texttt{static const} variables. This system is an efficient hierarchical structure flexible enough to encode the language idioms of C++.

\subsubsection{Conditional Compilation}

The structure of programs has one final issue before it can be considered an effective C++ substitute: the issue of conditional compilation. C++ is a two-tier language; the C++ program is first preprocessed using the wholly independent C preprocessor language which delivers the compilable pure C++-code. One particularly useful feature of the preprocessor is to allow a program to contain two sets of code which compiled together would result in error (possibly structural), but when compiling either one results in a correctly functioning program. Since the present proposal requires the entire program to be correct at all times, it is not clear how this could be addressed not least in terms of concept type-correctness.

The solution here revolves around giving a node the ability to redirect a dereference away from itself to one if its children. This idea is named node-rerouting. Thus the program remains strictly hierarchical but nodes may elect---except for archival and display---to disappear, effectively hiding behind one of their children. Thus there is the \textit{actual} structure (analogous to the whole C++ program) and the \textit{normal} structure (analogous to the pre-processed C++ program).

In addition to addressing the problem of conditional-compilation language constructs, this addresses a more serious problem of including parameters within literal program fragments\footnote{program fragments are often used in program transformation functions (e.g. defining a new concept in terms of an existing one or writing refactoring algorithms) and save the programmer having to painstakingly construct a required program fragment manually}. Consider a literal program fragment of the form \texttt{T t; S(\&t); cout<<t;}. The programmer wishes to parameterise on \texttt{T} \& \texttt{S} deciding the actual type and function according to factors available only at run-time. The nodes representing \texttt{T}/\texttt{S} would obviously be derived from a specialised type (perhaps \texttt{FragmentParameter}) of no relation to the types allowed at those positions i.e. \texttt{TypeReference} or \texttt{ValueReference}). With rerouting, the node could be \texttt{FragmentParameter}, but for the purposes of gaining structural acceptance from its parent could redirect itself to a hypothetical child of the correct concept class.
 
\begin{figure*}[t]
\centering
\includegraphics[scale=0.8]{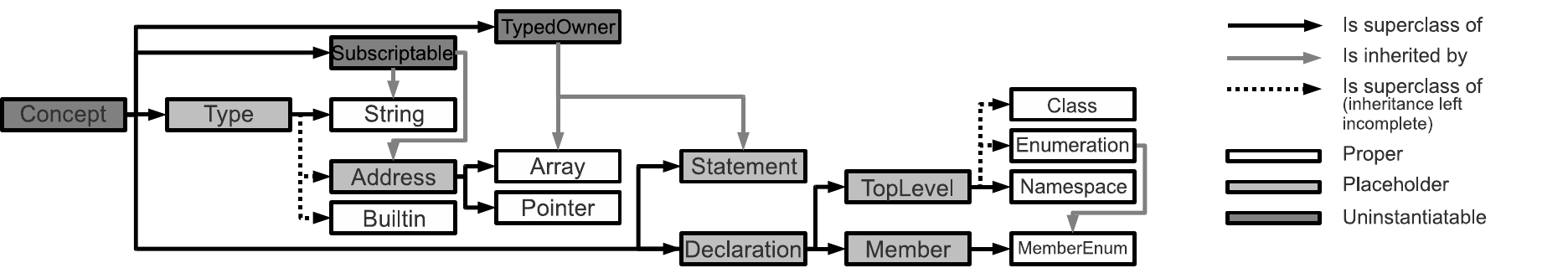}
\caption{A sample of the C++-like abstract language tree. Using multiple inheritance, the concept \texttt{Enumeration} which is accepted as a valid type of child for nodes of, \eg the \texttt{Namespace} concept, may be inherited into \texttt{MemberEnumeration} whose superclass is \texttt{Member}, providing extensibility and code sharing. \textit{Namespace} would not allow \texttt{MemberEnumeration} as an acceptable concept-type for its children as its strict (`is-a') ancestry (starting with \texttt{Member}) is dismissed by \texttt{Namespace}, but a hypothetical \texttt{Class} concept whose children must be \texttt{Member} concepts would accept it.}
\label{fig:cpphier}
\end{figure*}

\subsection{Intermediate Structure}
\label{sec:intermediatestructure}

During the user's definition of their program with a necessarily serialised input stream, there is usually not enough information to construct a well-defined and, more importantly, structurally correct program. If, for example, the user were to enter the simple expression \texttt{1+1}, then prior to making the third keystroke it is not clear exactly what structure the program should have\footnote{\cite{birnbaum2005} provides a more extensive critique}. As the system must maintain structural correctness at all times (if, for nothing else, to present a usable display of the program), this presents a clear difficulty. Language workbenches such as Domain Workbench have circumvented this problem by requiring the user to enter program fragments as text, thereby evading structurally-malformed intermediate stages of entry of the program.

For the present system such a fallback was considered unacceptable. As such, concepts were introduced that represent nothing more than indecision; there are called \textit{placeholders}. By filling with placeholders the child slots of nodes who would otherwise by structurally malformed, they allow an incomplete program to be structurally correct but semantically indeterminate.

Rather than have a single `wild-card' type of placeholder, placeholders are properly typed and are determined automatically as the most specialised common ancestor of all concepts allowed at the position to be filled. So, immediately prior to third stroke of the previous example, the structure would be an addition operation parenting two children, one a literal of \texttt{1}, the other a node of type \texttt{Expression} (a placeholder concept).

Placeholders, being concepts, may therefore have UI events routed to them. Given this, the placeholder mechanism allows an effective modular and extensible input mechanism to be built. Such UI events may be interpreted by the placeholder itself or (more likely) routed automatically to each of its derived concepts until one accepts the input and acts on it. This compartmentalises the input mechanism and helps send event information only to routines that might use it.

There are therefore three broad types of concept; the \textit{placeholders} described above, \textit{notional} concepts described previously, and the other ``complete'' concepts which we will call \textit{proper}. If a concept is notional (such as the \texttt{Concept} class itself), it is not instantiatable; it is used to define an abstract interface or shared behaviour. \Eg \texttt{TypedOwner} represents the ability of a node to place value-type restrictions on its children.

Proper concepts are well-defined and concrete (\eg \texttt{Invocation} represents a function invocation); their nodes typically have associated data or children and specify I/O behaviour. Placeholders, however, exist only to guarantee structural correctness part way through program entry (\eg \texttt{Statement} represents an undefined program statement, as you might expect when you press the Enter key in a function block) and as such their presence normally guarantees static-semantic error.

\subsection{Concept Inheritance}
\label{sec:inheritance}

There is a grave issue regarding the design of a \textit{concept} as expressed so far: As previously mentioned, it is useful for concepts to able to define structural requirements by limiting the concepts allowed for their nodes' children to a specific set. Only concepts derived from a member of, or actually within, this set would be allowed as children, others would not. This mechanism is provided effectively by the classic object-orientated `is-a' function, realised in C++ through, \eg \texttt{dynamic\_cast}.

There are, however, clear examples where inheriting from multiple classes is effectively a requirement. Consider the C++ concept \texttt{MemberVariable}. It should clearly be inherited from the \texttt{Member} concept (it makes sense to dictate that the children of a node of type \texttt{Class} should be \texttt{Member}s). However, there is a strong argument for inheritance also from \texttt{Variable}; \texttt{Variable} would have a considerable amount of code that might be useful, may itself inherit from concepts that make sense for \texttt{MemberVariable} (\eg \texttt{Typed}), and may be referenced by other nodes in the program, \eg \texttt{VariableReference}, a concept that represents a variable in use.

There are a number of solutions to this; one is to ignore the merits of using C++ class inheritance and use a run-time mechanism for determining such attributes instead. This would however result in no compile-time checks and potentially unsafe object casts. Another is to follow the traditional route and inherit from \texttt{Member} but use composition for \texttt{VariableReference}. This is ineffective here as it loses referential transparency; the \texttt{VariableReference} node would not share the attributes of its owner (such as children/parent). Any references to it expecting a node within the program would be broken.

The solution I provide here is to inherit from both, but retain a difference between \textit{superclass} and merely \textit{inherited by}.For \texttt{MemberVariable}, \texttt{Member} is named as the superclass, but it also derives from \texttt{Variable}. Thus there is strong fulfilment (or the `is-a' relationship) and weak fulfilment (or the `inherits' relationship). Weak fulfilment is equivalent to normal (multiple-)inheritance in C++, and is used for normal reference (or pointer) interface semantics. Strong fulfilment puts a single-inheritance hierarchy over a subset of the multiple-inheritance tree, and is used for determining node-ownership semantics.

Clearly, each node must have exactly one set of child nodes. With the base Concept class being defined as the class common to all types of nodes, it manages these relationships. Thus Concepts, if they use multiple inheritance must always inherit virtually, safeguarding the fact that each node has stored only one set of children.

Therefore, concept-derived classes:

\begin{itemize}
\item are stated as being either \textit{proper}, a \textit{placeholder} or \textit{notional};
\item must inherit at least one concept class or \texttt{Concept} itself;
\item must always inherit from concept classes virtually;
\item must nominate a single `superclass' from the set of inherited concept classes;
\item may be queried at run-time and compile-time on the identity of `superclass', membership of the power-set of `superclass' and concept type.
\end{itemize}

\subsection{Editing}

It is not immediately clear how the editor actually manages the editing interface to the user, while remaining extensible and modular. To address this, the present solution involves two complementary systems. The first, \textit{actions}, are general program transformations. The second, \textit{editable concepts} are concepts which may enter a special ``editing'' mode whereby it temporarily receives all input events and may represent itself differently.

\subsubsection{Actions}

Actions are program transformations, required to be reversible, involving nodes of a particular concept: user-directed structural program manipulation is done through a sequence of actions. Concepts register their repertoire of actions at link-time. Through the GUI editor, an action may have a user-configurable mapping to a concrete UI event, such as key sequences, GUI menu items, buttons, mouse gestures \&c., though each action is typically provided with a sensible default (often a key-stroke or sequence).

Actions are typically used for creation and placement of a node in the program, but could also be for property manipulation or to replace an existing proper node with a one of another type. All actions are reversible, which serves two purposes: it allows a comprehensive ``undo'' mechanism which, I assert, is an important adjunct since there is no immediate equivalent of the text-editor's backspace key. It also allows partially colliding multi-key sequences to be allowed without sacrificing timeliness of response. \Eg The two input sequences `\texttt{i+6}' and `\texttt{i++}' both begin with the same input sequence; determinism and timeliness of response require the program model to be in an ultimately incorrect intermediate state for at least one of the expressions. In this system's case, the user would see a program of the form \texttt{i + ?} following the first two key strokes---acceptable for the former, but not the latter, program. A third keystroke of `\texttt{+}', in forming a recognised stroke-sequence (`\texttt{++}') would reverse the action of the second stroke (forming a binary addition operation) and conduct the appropriate action (forming a unary increment operation). This system thus places no dependence relationship on the two operation concepts, a requirement for the abstract input system proposed.

\subsubsection{Editables}

Actions, as previously described, specify the primary mechanism for entering and editing the structure of a program. However, a small minority of concepts, generally those with properties, require considerably more involved editing systems that can comfortably be provided with such a simple mechanism. For this reason, the notion of \texttt{Editable} was created. An editable node is one that is able to enter an ``editing'' state (typically mapped to an otherwise unrecognised, non-control, input key press) whereby normal input and display mechanisms are delegated to an object particularly suited for such a datatype. Examples of such concepts include \texttt{ReferencedValue} and \texttt{IntegerLiteral}. Several sub-classes of \texttt{Editable} exist within this extension including \texttt{IntegerEditable} and the more generic \texttt{CompletionListEditable}. The editor extension recognises nodes of these types and induces the correct action upon an attempt to edit such a node. This may be to delegate the task to the currently active display extension, or simply to invoke a GUI dialogue.

Editing is not limited to nodes with properties or even proper nodes: the placeholder node \texttt{Statement} may, for example, be editable. In this case, the result of the edit is not to change the properties of the node, but rather to replace the node with some other, less general, node (or nodes). A simple system of registration is employed allowing arbitrary concepts may register potential names or keywords with such placeholders so that they may be recognised and acted upon correctly if entered. Examples of such keywords include type identifiers (whereby entry introduces a local variable declaration), variable/function identifiers (introducing a reference to that value) and C++ keywords (\eg \texttt{BoolLiteral} registers the keyword \textbf{true} and \texttt{ForStatement} registers \textbf{for}, both of which introduce the corresponding concepts).

\subsection{Program Semantics and Execution}
\label{sec:semantics}

The exact meaning of a program is defined by its actions on execution. In this case, the execution is determined solely by the node at the root of the program tree. The type of this node must be derived from a notional concept called \texttt{Program}. This notion defines a routine and/or a number of commands that must be executed on the system in order for the program to be executed. Notably, though other concepts in the abstract language may elect to describe their semantics in some way (such as a substitution into elementary C++ code or a structure of more basic nodes), it is ultimately the root \texttt{Program} node that determines the semantics of the given program.

In the case of the C++ abstract language, the steps to execution involve composing the pure C++ code for the program, compiling each module of the program, linking the resultant object code and finally running the executable. In the case of an interpreted language, \eg Scheme, the compilation/linking step would not be needed. In principle, an abstract code-building back-end could be used to make a given abstract language not only compiler-independent but implementation-language-independent through implementing the \texttt{Program} concept effectively as a `back-end' for each language to be supported. 

\subsubsection{Program Transformations}

Though each node may define itself in a textual C++ representation, this, in general, is not enough for true language-orientated programming: new concepts must be able to define their semantics as a program transformation to arbitrary, more primative, concepts. Unlike other approaches, even this mechanism is not provided in the core of the system, but rather through the notional concept \texttt{Composite} and the C++ abstract language's specialised \texttt{Program} concept:

As a step prior to the formation of the pure C++ code for the program, the program is first transformed. A \texttt{Composite} notional concept handles the process and it is here that concepts of otherwise undefined semantics regarding the basic C++ \texttt{Program} may define their meaning. The transformation process is, for this initial design, trivial but general; a node may inherit from \texttt{Composite} and will be called upon to transform itself. Arbitrary code is allowed when defining the transformation. The process of transforming the program continues through the tree and repeats as long as any \texttt{Composite} nodes remain. Though the order of node transforms is arbitrary, nodes are free to inspect the program tree and delay their transformation or otherwise coordinate transforms with other nodes. This is a somewhat brute-force approach to program transformations, and may be built upon at a later date with more refined techniques such as those described by \cite{bravenboer2008}.

\begin{figure}[H]
\label{fig:cyclic}
\begin{center}
\includegraphics[width=8.1cm]{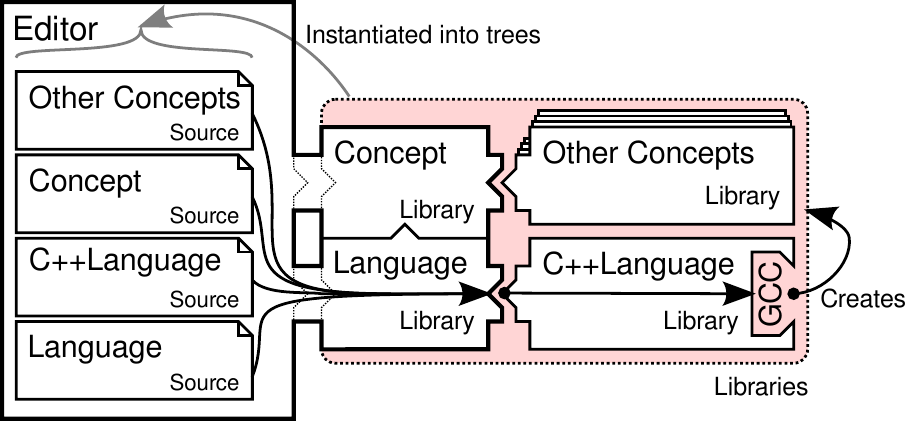}
\end{center}
\caption{Martta's self-referential nature---the arrow shows the direction of the initial bootstrap stage.}
\end{figure}

\section{Implementation}

In terms of implementation, the project is C++-based, cross-platform and is open-source with an open-development model. Cross-platform, open-source technologies were also co-opted to speed development: Apple's \textit{WebKit} was used at the primary engine for the display (utilised from the WebView GUI extension). Nokia's \textit{Qt} was used as the toolkit for the GUI. Finally, KitWare's \textit{GCC-XML} was used to analyse existing C/C++ code in order to translate them into a Martta program-fragment.

Though by no means complete, Martta fast approaches the functionality of a full language workbench. The code-base of my initial implementation of this design is just over 20,000 lines of C++. This is an order of magnitude smaller than \eg MPS (over 170,000 lines of Java), and though it would be foolish to suggest the two were directly comparable, I would nonetheless take it as evidence to suggest that the present approach is effective.

\subsection{Display Example}
\label{sec:DE}

\begin{figure}[H]
\begin{center}
\includegraphics[width=8.2cm]{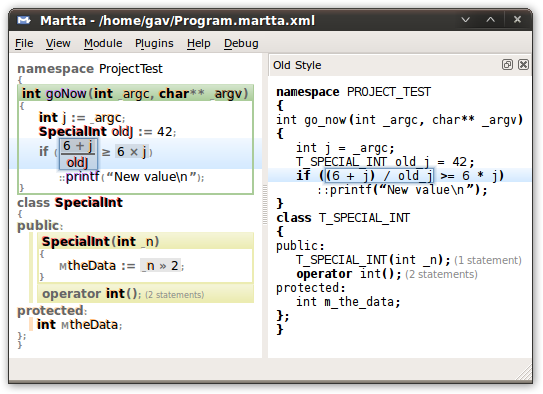}
\end{center}
\caption{A screenshot of Martta demonstrating two views onto the same program.}
\label{fig:screenshot}
\end{figure}

Figure \ref{fig:screenshot} demonstrates some of the power of language workbenches. Here, the same simple program has two differing views. The two views are configured independently, and in this case, differently. Importantly, the program model, and thus the serialised data that can be saved to a file or versioned in a revision control system remains unchanged.

On the left of the window is the default view for Martta programs. This uses camel-cased identifiers and background tones, rather than parentheses, to denote the expression hierarchy. On the right side is a more traditional view using same-case underscore-based identifiers and parenthesised operations to make clear any expression trees whose C++ form would require parentheses. 

Modern, partially semantic-aware text-editor features such as \textit{code-folding}, or the toggleable hiding of a function's body, are trivially implemented using the WebView's JavaScript engine. In this case the bodies of the two methods of the \texttt{SpecialInt} class are hidden from right view with on the cast operator's body hidden in the left. Double-clicking on the title area or pressing the \texttt{\{}/\texttt{\}} keys shows/hides the body.

\subsection{Input Example}
\label{sec:IE}

\begin{figure}[H]
\begin{center}
\includegraphics[width=5.7cm]{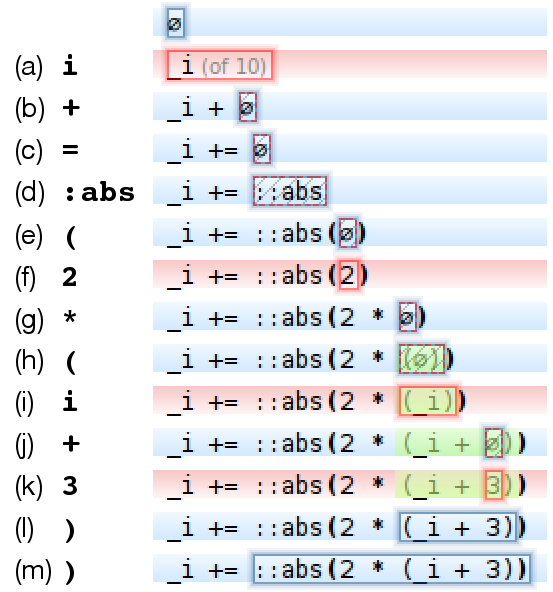}
\end{center}
\caption{14 stages of entry of the expression \texttt{i+=:abs(2*(i+3))}.}
\label{fig:inputstages}
\end{figure}

Figure \ref{fig:inputstages} demonstrates the visual appearance of the editor through the entry of a typical expression; \texttt{i+=:abs(2*(i+3))}. The evolving state of the program model should be made clearer through the automatic and inevitable display of placeholder concepts.

(b) and (c) demonstrate a multi-character input sequence; in this case the operator wanted is \texttt{+=}, which requires first the entry of the \texttt{+} operator. The editor's key entry system maintains a buffer of the input key strokes and gives priority to actions bound to longer keystroke sequences; if one is found, it reverses the effects of actions that were effected for earlier subsections of the keystroke sequence before executing the better action. In this case, with the \texttt{=} stroke in (c), the previous step's program augmentation is undone and the action to insert a new \texttt{+=} operation into the program is executed.

(b - e), (g), (h) and (j) demonstrate incomplete expressions; properly defined as a structurally correct, but semantically incorrect, model. The offending concept has a red dashed-line and grey crossings through it. (d) is the most interesting of these, with the editor determining a type mismatch between the function value \texttt{abs} and the expected \texttt{int} value. This is corrected in the following step where the \texttt{abs} value is usurped and adopted by an Invocation concept. A full explanation of the problem along with automatic solutions may be given off-screen.

Normally, expression entry, through the \texttt{Operation} concept, will follow normal C++ rules of precedence and associativity, changing the expression tree as is required to follow these rules. This mechanism can be altered by tagging certain concept positions: concepts at such positions appear green and parenthesised, and cannot have their program model altered implicitly. To tag a concept position in such a way, the parenthesis character is used in much the same way as in C++. (h - k) demonstrate the use of the tagging mechanism.

It is important to note that parentheses are not concepts and have no presence in the program model. In this view, the minimum number of parentheses are displayed such that the displayed code, when read as C++, is unambiguous and accurately reflects the program. It is key to note that although to the user, the parentheses around \texttt{\_i + 3} in (k) and in (l) appear identical (intentionally), they actually arise from completely different parts of the system. In (k) they are a temporary visual aid to denote a tagged concept position and last only as long as the expression entry. In (l) they are a display mechanism to more accurately reflect the entered program model. That, in the normal case, one disappears when the other initially appears is per design, however either can be separately disabled or otherwise customised at will.

\subsection{Proposed Directions}

\subsubsection{Revision Control}

Revision control has become common feature in integrated development environments, allowing the user to add/remove files view differences between revisions commit changes to a repository (perhaps with a commit string). In general, structural programming environments and language workbenches have no intrinsic serialised language. Thus no file format is imposed nor is there any need to store programs primarily in files for use with other components of the development stack. As such I would propose to remove from the editor the act of traditional file-saving, instead providing a repository-commit function. 

Most programmers save their work locally much more frequently than committing to a repository\footnote{though this is slowly changing as a new generation of distributed RCSs become commonplace. See \eg Git, Bazaar, Arch, Mercurial}. Each save point will likely have some rationale behind the changes made; it might represent a bug potentially fixed, some debug information added or a (partial) feature implementation \&c., and as such should deserves to be revision commit and not a blind-save. Co-ordination with a lightweight tasks infrastructure (discussed next) and judicious use of tags/branches and hierarchical levels of repositories would minimise interruption to the workflow while providing a fine-grained record of changes made to the code-base.

A guarantee of versioning functionality within a structural editor gives an interactive semantic history of a codebase, making practical certain features. In terms of project management, it would allow a project manager or programming lead to get an informative overview about where in the codebase developers concentrate their efforts, which developers work closely on which parts of the code base and recurring collisions over shared code. This knowledge could be used to improve division of labour, scheduling and general team management.

For a structural editor, and especially a language workbench, however, revision control presents an interesting problem---that of representing a set of changes to a program without loss of generality or effectiveness of representation. Further work is needed in this area.

\subsubsection{Problem, Unit-tests, Bugs and To-dos}

An abstract infrastructure for the reporting and managing of tasks regarding a program would be an interesting adjunct to a language workbench such as Martta. Complete semantic awareness at the UI level allows some blurring of the distinction between a program's structural semantic correctness and a program's conceptual semantic correctness. For example, as shown above it is trivial to determine, list, mark-up and, in some cases, solve, in real-time, what would be semantic compilation issues such as expression type-correctness. Some current editors offer similar functionality, albeit with some reduction in accuracy, precision and/or timeliness.

The traditional \texttt{TODO:}, \texttt{BUG:} and \texttt{FIXME:} notes often found in program code could easily be integrated well into the infrastructure. A specific concept, in effect a semantic comment, could represent such a note. A unit-test subsystem with access to the same infrastructure would allow a slightly higher-level of issue reporting, albeit less timely. A bug-tracking subsystem, perhaps with some back-end connection to an online bug-tracking database, would allow a further level of issue reporting, less timely and precise still.

This issue listing/annotation system could similarly be integrated with the revision control system above to automatically note issues fixed (or broken) on given commits.

\subsubsection{Metasyntactic Variables}

When prototyping code, the names of values is often of little importance. Programmers typically resort to using common placeholder variable names among the most common are \texttt{foo} and \texttt{bar}. Martta already provides a naming system for anonymous identifiers; identifiers with an empty text label are given a name of the form \texttt{\_anon1234567}. Allowing the programmer to define and customise these placeholder names such that they come from a list of his choosing would reduce the length of such names and eliminate unnecessary key strokes.

\subsubsection{Invocation Expansion}

A relatively common occurrence when working with a code base is that the programmer comes across a function invocation and wishes to see its implementation in order to better understand its behaviour and limitations. Traditionally, this would mean locating the implementation source code file containing the function, displaying that file at the correct point, and eventually navigating back to the file containing the invocation. Modern editors, \eg Trolltech's Qt Creator (described by \cite{nokia2009}), are able to ease this process considerably by providing search and navigation functionality for language entities. However, ultimately, there exists a required mental discontinuity of thought in navigating from invocation to implementation that conceptually should not be.

The temporary expansion of (statically-dispatched) invocations to the semantically-equivalent inline code mitigates this discontinuity. It permits the programmer to selectively ignore the imposed language-abstraction of functions and its invocation/implementation duality, instead to concentrate on the meaning of the invocation in terms of the execution path. The fact that the program is stored internally and rendered to display makes a crude version of this functionality trivial. More advanced program analysis (such as determining referential transparency and solving static arithmetic) would be required to optimise the expansion's appearance.

A simple example might be the program represented by the Fibonacci sequence generator:

\begin{verbatim}
uint fib(uint n) {
  if (n <= 1) return n;
  return fib(n-1)+fib(n-2);
}
\end{verbatim}

Invoking the \textit{invocation expansion} functionality of the GUI over the first \texttt{fib} invocation would result in the display of a program represented by:

\begin{verbatim}
uint fib(uint n) {
  if (n <= 1) return n;
  uint _arg = n-1;
  if (_arg <= 1)
    return _arg + fib(n-2);
  return fib(_arg-1)+fib(_arg-2) + fib(n-2);
}
\end{verbatim}

With basic program analysis techniques, \texttt{\_arg} could be substituted for the argument, giving, with minor simplifications:

\begin{verbatim}
uint fib(uint n) {
  if (n <= 1) return n;
  if (n <= 2)
    return n - 1 + fib(n - 2);
  return 2 * fib(n - 2) + fib(n - 3);
}
\end{verbatim}
% return 2*fib(n-2)+fib(n-3);

\subsubsection{Auto Prototyping}

Unless exceptionally foresighted, most programmers will, while implementing a routine, realise that another (usually lower-level) subroutine needs to be written. Assuming she does not immediately drop the task at hand and start coding the lower-level subroutine, two options are available: write the interface for the new subroutine before finishing the routine, or, finish the routine (using the hypothetical new subroutine) then write its interface. The former, since it keeps the program in a semantically valid state, is perhaps favourable, but does require the programmer to break with his task at hand for a short period.

In a language workbench it becomes largely trivial to have the editor automatically determine that a particular symbol (perhaps a variable or method) does not exist and query the user if it should be created. Furthermore, contextual information (such as the expected type and its invocation arguments) can be used to estimate a full interface to the method. Combined with invocation expansion, above, the subroutine could be coded conceptually \textit{inline} allowing all to be done without the programmer resorting to code navigation or otherwise breaking their concentration.

\subsubsection{Inline Program Analysis}

Program analysis, particularly in C++, is traditionally swamped with the prerequisite that the program's source code must first be turned into an accurate representation of the program's meaning (an abstract syntax tree, or some other abstract intermediate-level structure). This is a particularly difficult task---there are several commercial projects who specialise in creating proprietary software for this. \cite{sim2002} lists and evaluates several of the contenders in terms of accuracy and robustness, concluding none delivered perfect results.

If, however, the program editor stores the program in terms of its meaning, the act of program analysis bypasses this step completely. Interactivity delivers the possibility for real-time indication and solving of troublesome program fragments. Interactive expression solving and simplification within the editor becomes practical, if not trivial. The annotation of attributes of blocks of code, \eg referential transparency, \textit{const}-ness of a class method, a structural dependency overview on block of code (program entities used, potential changes made \&c.) are trivially made. Key to all of this is that the functionality is up-front in the editor and timely.

%Non-primitive concepts may need only to be defined in terms of their transformation; semantic stuff \& dependencies could be inferred.

\section{Conclusion}

In this introductory paper I have outlined a design of a C++-based language workbench, and discussed my cross-platform, open-source implementation of this design, \textit{Martta}. In so doing I have demonstrated it not only possible, but effective, for C++ to be used as the platform-language for a language workbench, hitherto implemented only with dynamic dispatch, run-time compiled programming environments such as Java and C\#. My basic implementation shows signs that the design, together with the use of modern middleware where possible, delivers a comparably small codebase given the functionality it delivers.

The design approach I have taken differs in several respects from those already proposed. Some differences, such as the judicious usage of multiple inheritance, make implementation in other languages impractical or impossible. Other differences, such as the tiered customisability of the display and the absolutist uncoupled approach to program input and display are not platform-language-specific. Two user-interface examples demonstrate the initial promise in this design that I have found through in my own informal user-interface testing.

Finally I have discussed several further directions this design, and language workbenches in general, could be taken.

\bibliography{Biblio}
\bibliographystyle{plainnat}
\end{multicols}

%\appendix
%\begin{landscape}
%\begin{figure}[ht]
%Appendix A: Base language graph.
%\centering
%\label{fig:langauge}
%\includegraphics[width=24cm]{language.pdf}
%\caption{Some of the key language components of the base language in Martta. Red/blue/back indicate notional/placeholder/proper nodes and solid/dashed edges represent strong/weak inheritance respectively.}
%\end{figure}
%\end{landscape}
\end{document}